\documentclass[oneside,a4paper,11pt,shownumbers]{article}
\usepackage{latexsym}
\usepackage{euscript}
\usepackage{epsfig,amsmath,amssymb}

\def\be{\begin{equation}}
\def\ee{\end{equation}}
\def\bea{\begin{eqnarray}}
\def\eea{\end{eqnarray}}

\long\def\symbolfootnote[#1]#2{\begingroup%
\def\thefootnote{\fnsymbol{footnote}}\footnote[#1]{#2}\endgroup} 

 \large\normalsize

\begin{document}

\begin{center}

{\Large \bf Cosmic strings interacting with dark strings}

\vspace*{7mm} {Betti Hartmann $^{a}$
\symbolfootnote[1]{E-mail:b.hartmann@jacobs-university.de} and
Farhad Arbabzadah $^{a}$
\symbolfootnote[2]{E-mail:f.arbabzadah@jacobs-university.de}}
\vspace*{.25cm}

${}^{a)}${\it School of Engineering and Science, Jacobs University Bremen, 28759 Bremen, Germany}\\

\vspace*{.3cm}
\today
\end{center}

\begin{abstract}
Motivated by astrophysical observations of excess electronic production
in the galaxy, new theoretical models of the dark matter sector have been
proposed in which the Standard model couples to the dark matter sector through
an attractive interaction.
The coupling of the Standard model to  ``dark strings'', which are solutions
of the low energy dark sector has been investigated recently.
Here, we discuss the interaction between dark strings and standard cosmic strings and show that
they can form bound states.
In the presence of the interaction term, a Bogomolny-Prasad-Sommerfield (BPS) bound exists that depends on the interaction parameter and we observe that the attractive interaction
between dark strings and cosmic strings is most efficient if the two strings
are identical.
Moreover, our model allows for dark string solution that can lower their energy by coupling to
an electromagnetic field.
We also investigate the gravitational properties of these solutions
and show that they become supermassive with a singular space-time for values of the gravitational
coupling larger as compared to the non-interacting case. Moreover, the deficit angle
of the solutions decreases with increasing interaction.

\end{abstract}

\section{Introduction}
There is strong observational evidence \cite{observation_dm} that approximately $22\%$ of the total energy density of the universe is in the form of dark matter. Up until now it is unclear what this dark matter should be made of. One of the favourite candidates are Weakly Interacting Massive Particles (WIMPs) which
arise in extensions of the Standard Model.
Recently, new theoretical models of the dark matter sector have been proposed \cite{dark}, in which
the Standard Model is coupled to the dark sector via an attractive interaction term.
These models have been motivated by new astrophysical observations \cite{observation_ee} which show an excess 
in electronic production in the galaxy. Depending on the experiment, the energy of these
excess electrons is between a few GeV and a few TeV.
One possible explanation for these observations is the annihilation of dark matter 
into electrons. Below the GeV scale, the interaction term in these models is basically of the form of a direct
coupling between the U(1) field strength tensor of the dark matter sector and the U(1) field strength tensor of
electromagnetism. The U(1) symmetry of the dark sector  has to be spontaneously broken, otherwise a ``dark photon'' background leading to observable consequences would exist.

Consequently, it has been shown that the dark sector can have string-like solutions, denominated ``dark strings'' 
and the observational consequences of the interaction of these dark strings with the Standard Model
have been discussed \cite{vachaspati}.

Topological defects are believed to have formed in the numerous phase transitions in the early
universe due to the Kibble mechanism \cite{topological_defects}.
While magnetic monopoles and domain walls, which result from the spontaneous
symmetry breaking of a spherical and parity symmetry, respectively, 
are catastrophic for the universe since they would overclose it, cosmic strings
are an acceptable remnant from the early universe. These objects 
form whenever an axial symmetry gets spontaneously broken and (due to topological arguments)
are either infinitely long or exist in the form of cosmic string loops. Numerical
simulations of the evolution of cosmic string networks have shown that
these reach a scaling solution, i.e. their contribution to the total energy density
of the universe becomes constant at some stage. The main mechanism that allows
cosmic string networks to reach this scaling solution is the formation
of cosmic string loops due to self-intersection and the consequent decay of these loops
under the emission of gravitational radiation.

For some time, cosmic strings were believed to be responsible for the structure
formation in the universe. New Cosmic Microwave background (CMB) data, however, clearly
shows that the theoretical power spectrum associated to Cosmic strings
is in stark contrast to the observed power spectrum. However, there has been
a recent revival of cosmic strings since it is now believed that cosmic  strings
might be linked to the fundamental strings of string theory \cite{polchinski}.

While perturbative fundamental strings were excluded to be observable on cosmic scales
for many reasons \cite{witten}, there are now new theories containing
extra dimensions, so-called brane world model, that allow to lower the fundamental
Planck scale down to the TeV scale. This and the observation that
cosmic strings generically form at the end of inflation in inflationary models
resulting from String Theory \cite{braneinflation} and Supersymmetric Grand Unified Theories \cite{susyguts}
has boosted the interest in comic string solutions again.
The interaction of cosmic strings has been investigated in the context of field theoretical
models describing bound systems of D- and F-strings, so-called p-q-strings \cite{saffin,hu}.

Here we study the interaction of cosmic strings with dark string solutions.
On a field theoretical level, the model is similar to the one used in \cite{saffin,hu}, however
in this paper, the interaction between the strings is mediated via the gauge fields (and gravity), while
the strings in \cite{saffin,hu} interact via a potential (and gravity).
While the dark sector in our model is an Abelian Higgs model, we should in principle couple
the corresponding solutions to the electromagnetic field of the Standard model. Here, we assume the U(1) symmetry
to be spontaneously broken and the corresponding Abelian-Higgs model to possess cosmic
string solutions. In fact, we  employ the strategy that
the Standard model (and its semilocal brother) have string-like solutions \cite{semilocal}
which share many features with the solutions of the U(1) toy model. Solutions
in this U(1) Abelian-Higgs model have been first discussed in \cite{no}.
These solutions have a magnetic field with a quantized magnetic flux and
are topolgical solitons in the sense that they have a topological charge associated
to them. When the
gauge boson mass is equal to the Higgs boson mass in this theory, the string
solutions fulfill an energy bound, the Bogomolny-Prasad-Sommerfield (BPS) bound \cite{bogo}, such that
the energy per unit length is directly proportional to the topological charge.
These solutions have been discussed extensively and their gravitational
properties have also been investigated \cite{clv,bl,linet}. The main feature
of the space-time around a cosmic string is that it is locally flat, but globally
possesses a deficit angle $\delta$ that is directly proportional to the energy per unit length of the solutions $\mu$: $\delta=8\pi G \mu$, where $G$ is Newton's constant.
This leads e.g. to gravitational lensing effects that should make 
it possible to detect cosmic strings in the universe.
Interestingly, globally regular gravitating
strings exist only as long as the solutions are not too massive. If they become too massive
(or the gravitational coupling becomes too large) the deficit angle is larger than $2\pi$ and the
space-time is singular \cite{supermassive}. It has been noted \cite{bl,linet} that solutions
that are BPS in flat space-time are also BPS is curved space-time, i.e. fulfill the same
energy bound.

Our paper is organized as follows: in Section 2, we give the model, the equations and discuss
the BPS bound. In Section 3, we present our numerical results and Section 4 contains our conclusions and an outlook.

\section{The model}
We study the interaction of a U(1) Abelian-Higgs field model, which has cosmic string solutions,
with the low energy dark sector, which is also a U(1) Abelian-Higgs model in flat and
curved space-time, respectively.
Note that the U(1) model is a toy model here for standard model-like theories with gauge
group SU(2)$\times$ U(1), which
also contain string solutions. Examples would be semilocal strings in
the SU(2)$_{global}$ $\times$ U(1) model and
electroweak strings in the SU(2)$_{local}$ $\times$ U(1) model \cite{semilocal}, respectively.
We believe that our toy model captures the qualitative features of these theories.

The model we are studying is given by the following action:
\begin{equation}
\label{action}
S=\int d^4 x \sqrt{-g} \left( \frac{1}{16\pi G} R + {\cal L}_{m} \right)
\end{equation}
where $R$ is the Ricci scalar and $G$ denotes Newton's constant. The matter Lagrangian
${\cal L}_{m}$ reads:
\begin{equation}
{\cal L}_{m}=D_{\mu} \phi (D^{\mu} \phi)^*-\frac{1}{4} F_{\mu\nu} F^{\mu\nu}
+D_{\mu} \xi (D^{\mu} \xi)^*-\frac{1}{4} H_{\mu\nu} H^{\mu\nu}
-V(\phi,\xi) + \frac{\varepsilon}{2} F_{\mu\nu}H^{\mu\nu}
\end{equation} 
with the covariant derivatives $D_\mu\phi=\nabla_{\mu}\phi-ie_1 A_{\mu}\phi$,
$D_\mu\xi=\nabla_{\mu}\xi-ie_2 a_{\mu}\xi$
and the
field strength tensors $F_{\mu\nu}=\partial_\mu A_\nu-\partial_\nu A_\mu$, 
$H_{\mu\nu}=\partial_\mu a_\nu-\partial_\nu a_\mu$  of the two U(1) gauge potential $A_{\mu}$, $a_{\mu}$ with coupling constants $e_1$
and $e_2$.
$\phi$ and $\xi$ are complex scalar fields (Higgs fields) with potential
\be
V(\phi,\xi)=\frac{\lambda_1}{4}\left(\phi\phi^*-\eta^2_1\right)^2
+\frac{\lambda_2}{4}\left(\xi\xi^*-\eta^2_2\right)^2
\ee
The term proportional to $\varepsilon$ is the interaction term \cite{vachaspati}.
To be compatible with observations, $\varepsilon$ should be on the order of $10^{-3}$. 

In the following, we associate the dark strings to the fields $A_{\mu}$ and $\phi$,
while the standard cosmic strings are described by the fields $a_{\mu}$ and $\xi$.
The Higgs fields have masses $M_{H,i}=\sqrt{\lambda_i} \eta_i$, 
while the gauge boson masses are $M_{W,i}=\sqrt{2}e_i \eta_i$, $i=1,2$.

\subsection{The Ansatz}

In the following we shall analyse  the system of coupled
differential equations associated
with the gravitationally coupled system described above. This system will contain the Euler-Lagrange equations
for the matter fields and the Einstein equations for the metric fields. In order to do that,
let us write down the matter and gravitational fields as shown below.
The most general, cylindrically symmetric line element invariant under boosts
along the $z-$direction is:
\begin{equation}
ds^2=N^2(\rho)dt^2-d\rho^2-L^2(\rho)d\varphi^2-N^2(\rho)dz^2 \ .
\end{equation}
The non-vanishing components of the Ricci tensor $R_{\mu}^{\nu}$ then read \cite{clv}:
\begin{eqnarray}
R_0^0=-\frac{(LNN')'}{N^2 L} \ \ , \ \ R_{\rho}^{\rho} = -\frac{2N''}{N}-\frac{L''}{L} \ \ \ , \ \ R_{\varphi}^{\varphi}= -\frac{(N^2 L')'}{N^2 L} \ \ \ , \ \ \ R_z^z=R_0^0
\end{eqnarray}
where the prime denotes the derivative with respect to $\rho$. 

For the matter and gauge fields, we have \cite{no}:
\begin{equation}
\phi(\rho,\varphi)=\eta_1 h(\rho)e^{i n\varphi} \ \ , \ \  \xi(\rho,\varphi)=\eta_1 f(\rho)e^{i m\varphi} \
\end{equation}
\begin{equation}
A_{\mu}dx^{\mu}=\frac {1}{e_1}(n-P(\rho)) d\varphi \ \ , \ \ a_{\mu}dx^{\mu}=\frac {1}{e_2}(m-R(\rho)) d\varphi \ .
\end{equation}
$n$ and $m$ are integers indexing the vorticity of the two Higgs fields  around the $z-$axis.

\subsection{Equations of motion}
We define the following dimensionless variable and function:

\begin{equation}
x=e_1\eta_1 \rho \ \ \ , \ \ \ L(x)= \eta_1 e_1 L(\rho) \ .
\end{equation}

Then, the total Lagrangian ${\cal L}_m \rightarrow {\cal L}_m/(\eta_1^4 e_1^2)$ depends only on the following dimensionless coupling constants

\begin{equation}
\gamma=8\pi G\eta_1^2 \ \ ,  \ \ 
\beta_i=\frac{2M_{H,i}^2}{ M_{W,1}^2}\frac{\eta_1^2}{\eta_i^2}=\frac{\lambda_i}{e_1^2} \ , \ i=1,2   \ ,
\end{equation}
 and the dimensionless ratios of the coupling
constants and vacuum expectation values, respectively
\begin{equation}
g=\frac{e_2}{e_1} \ \ , \  \ \
q=\frac{\eta_2}{\eta_1} \ \ . \ \ 
\end{equation}
Varying the action with respect to the matter fields and metric functions, we
obtain a system of 
six non-linear differential equations. The Euler-Lagrange equations for the matter field functions read:
\begin{equation}
\frac{(N^2Lh')'}{N^2L}=\frac{P^2 h}{L^2}+\frac{\beta_1}{2}(h^2-1)h
\end{equation}
\begin{equation}
\frac{(N^2Lf')'}{N^2L}=\frac{R^2f}{L^2}+\frac{\beta_2}{2}(f^2-q^2)f
\label{eqf}
\end{equation}
\begin{equation}
(1-\varepsilon^2)\frac{L}{N^2}\left(\frac{N^2P'}{L}\right)'=2 h^2 P + 2\varepsilon g R f^2 \ ,
\end{equation}
\begin{equation}
(1-\varepsilon^2)\frac{L}{N^2}\left(\frac{N^2R'}{L}\right)'=2 g^2 f^2 R + 2\varepsilon g P h^2 \ ,
\end{equation}
where the prime now and in the following denotes the derivative with respect to $x$.

We use the Einstein equations in the following form:
\begin{equation}
R_{\mu\nu}=-\gamma \left(T_{\mu\nu}-\frac{1}{2}g_{\mu\nu} T\right) \ \ , \ \ \mu,\nu=t,x,\varphi,z
\end{equation}
where $T$ is the trace of the energy momentum tensor $T=T^{\lambda}_{\lambda}$ and the
non-vanishing components of the energy-momentum tensor are (we use the notation
of \cite{clv}) with $i=1,2$:
\begin{eqnarray}
T_0^0 &=& e_s + e_v + e_w + u  \ \ , \ \ 
T_x^x = -e_s - e_v + e_w + u \nonumber \\
T_{\varphi}^{\varphi} &=&e_s - e_v - e_w + u \ \ , \ \ T_z^z =  T_0^0 
\end{eqnarray}
where
\begin{equation}
\label{contributions}
e_s= (h')^2 + (f')^2   \ \ \ , \ \ \ e_v = \frac{(P')^2}{2 L^2} + \frac{(R')^2}{2 g^2 L^2} -\frac{\varepsilon}{g}\frac{R'P'}{L^2} \ \ \ , \ \ \ e_w = \frac{h^2 P^2}{L^2} + \frac{R^2 f^2}{L^2} \end{equation}
and
\begin{eqnarray}
u & = & \frac{\beta_1}{4}\left(h^2-1\right)^2 + \frac{\beta_2}{4} \left(f^2-q^2\right)^2  \ . 
\end{eqnarray}

We then obtain
\begin{eqnarray}
\label{N1}
\frac{(LNN')'}{N^2 L}&=& \gamma\left[\frac{(P')^2}
{2 L^2}+ \frac{(R')^2}{2 g^2 L^2} -\frac{\varepsilon}{g}\frac{R'P'}{L^2}
-u\right] 
\end{eqnarray}
and:
\begin{eqnarray}
\label{N2}
\frac{(N^2L')'}{N^2L}&=&-\gamma\left[\frac{2 h^2 P^2}
{L^2}+\frac{2 R^2 f^2}{L^2}+\frac{(P')^2}{2 L^2}+
\frac{(R')^2}{2 g^2 L^2 } -\frac{\varepsilon}{g}\frac{R'P'}{L^2} + u\right]  
\end{eqnarray}

\subsection{Boundary conditions}
The requirement of regularity at the origin leads to the  following boundary 
conditions:
\begin{equation}
h(0)=0, \ f(0)=0 \ , \ P(0)=n \ , \ R(0)=m
\label{bc1}
\end{equation}
for the matter fields and 
\begin{equation}
\label{zero}
N(0)=1, \ N'(0)=0, \ L(0)=0 \ , \ L'(0)=1 \ .
\end{equation}
for the metric fields. 
The finiteness of the energy per unit length requires:
\begin{equation}
h(\infty)=1, \ f(\infty)=q \ , \ P(\infty)=0 \ , \ R(\infty)=0  \ .
\end{equation}

\subsection{Energy per unit length, magnetic fields and deficit angle}

We define as inertial energy per unit length  of a solution describing the interaction of a dark string
with winding $n$ and a cosmic string with winding $m$:
\begin{equation}
 \mu^{(n,m)}=\int \sqrt{-g_3} T^0_0 dx d\varphi
\end{equation}
where $g_3$ is the determinant of the $2+1$-dimensional space-time given by $(t,x,\varphi)$.
This then reads:
\begin{equation}
 \mu^{(n,m)}=2\pi\int_{0}^{\infty} NL \left(\varepsilon_s + \varepsilon_v + \varepsilon_w + u\right) \ dx
\end{equation}
Note that the string tension $T=\int \sqrt{-g_3} \ T^z_z dx d\varphi$ is equal to the energy per unit length.
In flat space-time ($G=0$) and $\varepsilon=0$, the energy per unit length of the solution is
given by
\begin{equation}
\mu^{(n,m)}=2\pi n \eta_1^2 g_1(\beta_1) + 2\pi m \eta_1^2   g_2(\beta_2) 
\end{equation}
where $g_1$ and $g_2$ are functions that depend only weakly on $\beta_1$ and $\beta_2$, respectively
with $g_1(2)=1$ and $g_2(2)=1$ in the BPS limit. In the following, we will set $\eta_1=1$ without loss of generality.

We can define the binding energy of an $n$-dark string with an $m$-cosmic string as
\begin{equation}
\label{binding}
 \mu^{(n,m)}_{bin}=\mu^{(n,m)}-n \mu^{(1,0)}-m \mu^{(0,1)}
\end{equation}

The magnetic fields associated to the solution can be given when noting that the gauge part of the Lagrangian
density can be rewritten as follows \cite{vachaspati}:
\begin{equation}
 -\frac{1}{4} F_{\mu\nu} F^{\mu\nu}-\frac{1}{4} H_{\mu\nu} H^{\mu\nu}+ \frac{\varepsilon}{2} F_{\mu\nu}H^{\mu\nu} \Rightarrow -\frac{1}{4} \tilde{F}_{\mu\nu}
\tilde{F}^{\mu\nu} -\frac{1}{4}(1-\varepsilon^2) H_{\mu\nu} H^{\mu\nu}
\end{equation}
with $\tilde{F}_{\mu\nu}=\partial_{\mu} \tilde{A}_{\nu}- \partial_{\nu} \tilde{A}_{\mu}$
where $\tilde{A}_{\mu}=A_{\mu}-\varepsilon a_{\mu}$.

The magnetic fields associated to the fields $\tilde{A}_{\mu}$ and $a_{\mu}$ have only a component in
$z$-direction. These components read: 
\begin{equation}
\label{magnetic}
 \tilde{B}_z(x)=\frac{-P'(x)+\frac{\varepsilon}{g} R'(x)}{e_1 L(x)} \ \ \ {\rm and} \ \ \   b_z(x)=-\sqrt{1-\varepsilon^2}\frac{R'(x)}{e_2 L(x)}  \ ,  
\end{equation}
respectively. The corresponding magnetic fluxes $\int d^2x \ B$ are
\begin{equation}
 \tilde{\Phi}= \frac{2\pi}{e_1}\left(n-\frac{\varepsilon}{g} m\right) \ \ {\rm and} \ \ 
\varphi=\sqrt{1-\varepsilon^2} \ \frac{2\pi m}{e_2} \ ,
\end{equation}
respectively. Obviously, these magnetic fluxes are not quantized for generic $\varepsilon$.

Finally, the deficit angle $\delta=8\pi G\mu$ of the solution can be read of directly from the derivative of the
metric function $L(x)$. For string-like solutions, the metric functions behave like $N(x\rightarrow \infty) \rightarrow c_1$ and $L(x\rightarrow \infty) \rightarrow c_2 x + c_3$, where $c_1$, $c_2$ and $c_3$ are constants.
The deficit angle is then given by:
\begin{equation}
\delta=2\pi (1-L'\vert_{x=\infty})=2\pi(1-c_2)  \ .
\end{equation}

\subsection{The Bogomolny-Prasad-Sommerfield (BPS) bound}
For $\varepsilon=0$, the model has BPS solutions which satisfy the energy
bound $\mu=2\pi n + 2\pi m$ both in flat \cite{bogo} as well as in curved space-time \cite{linet}
for $\beta_1=2$, $\beta_2=2$.

Here we will discuss the case $f= h$ ($\beta_1=\beta_2\equiv \beta$), $P=R$ ($n=m$, $g=1$) for $\varepsilon\neq 0$.
In flat space-time, the functions $N\equiv 1$ and $L\equiv x$ and the corresponding BPS equations read
\begin{equation}
 h'=\frac{Ph}{x} \ \ , \ \ (1-\varepsilon) \frac{P'}{x}=h^2-1
\end{equation}
The solutions then fulfill the energy bound $E=2\pi n + 2\pi m=4\pi n$ for 
\begin{equation}
 \beta=\frac{2}{1-\varepsilon}
\end{equation}

In curved space-time, i.e. for $\gamma\neq 0$, the above bound is  still a BPS bound.
The corresponding matter equations read
\begin{equation}
 h'=\frac{Ph}{L} \ \ , \ \ (1-\varepsilon) \frac{P'}{L}=h^2-1 \ .
\end{equation}
We have $e_v = u$ and $e_w=e_s$ such that $T_x^x=T_{\varphi}^{\varphi}=0$. Hence $N(x)\equiv 1$, while
$L$ fulfills the following equation:
\begin{equation}
\frac{L''}{L}=-2\gamma\left(\frac{2P^2h^2}{L^2} + \frac{1}{1-\varepsilon}(h^2-1)^2\right) 
\end{equation}
The deficit angle then reads
\begin{equation}
\label{deficitBPS}
\delta = 2\pi \gamma (n+m)   \ .
\end{equation}

\subsection{Dark strings coupled to an unbroken U(1) symmetry}
For $\beta_2=0$, $q=0$ (with $\eta_1$ finite), we have $f\equiv 0$ and  the corresponding U(1) symmetry
remains unbroken. For $\varepsilon=0$, $m=0$, we would then have $R\equiv 0$. However, for $\varepsilon \neq 0$, $m=0$ the
energy of the solution can be lowered by a non-vanishing derivative of $R$, i.e. $R'\neq 0$. This corresponds to
a non-vanishing magnetic field in $z$-direction. Due to the attractive nature of the interaction,
the dark string can thus lower its energy by coupling to a non-vanishing magnetic field inside the
string core. The corresponding
boundary conditions for $R$ then read:
\begin{equation}
\label{newbc}
 R(0)=0 \ \ \ {\rm and} \ \ \ R'\vert_{x=\infty}=0
\end{equation}
Note that we don't need $R$ but only $R'$ to vanish at infinity to fulfill the requirement of finiteness of the energy. (\ref{contributions}) then suggests
that the $e_{v}$ part of the energy density is minimized for
$R'=\varepsilon g P'$ inside the dark string core. Using the boundary conditions (\ref{bc1}), (\ref{newbc}) we find that $R=\varepsilon g (P-n)$. 
The energy of an $(n,0)$ string is thus smaller for $\varepsilon \neq 0$ than for $\varepsilon =0$. The same is, of course,
also true for the inverse case, where $n=0$ and $m\neq 0$ and $R$ replaced by $P$ in the boundary conditions (\ref{newbc}).

\section{Numerical results}
In all our numerical calculations, we have set $q=1$, $g=1$.
We have solved the set of coupled ordinary differential equations numerically using the ordinary differential
equation solver COLSYS \cite{colsys}. Numerical errors are typically on the order of $10^{-8}-10^{-9}$.

\subsection{$\gamma=0$}
To begin with, we present our results for cosmic-dark string systems in flat space-time $\gamma=0$. In this
case, we have $N\equiv 1$ and $L\equiv x$.

\begin{figure}[htbp]
\centering
    \includegraphics[width=13cm]{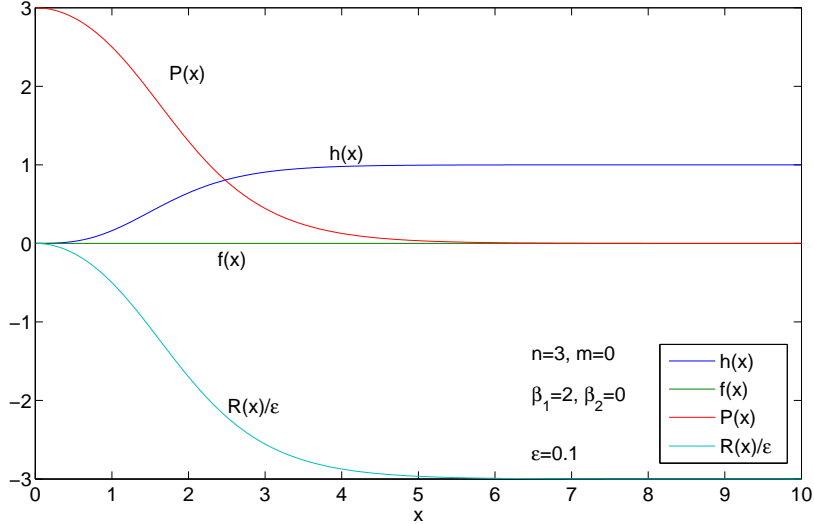} 
\caption{The functions of a  dark string solution for $n=3$, $m=0$, $\beta_1=2$, $\beta_2=0$ and $\varepsilon=0.1$. The fields associated to
the dark string are $P(x)$ and $h(x)$, while the scalar field of the unbroken U(1) symmetry
vanishes identically $f(x)\equiv 0$. The gauge field function $R(x)$ however is related
to the function $P(x)$ by $R(x)/\varepsilon=P(x)-n$.
 }
\label{fig1}
\end{figure}

We have first studied solutions with $m=0$, $\beta_2=0$. As discussed above, the field $f\equiv 0$, but
$R\neq 0$. A typical solutions of this type is shown in Fig.\ref{fig1}, where we present the functions
$P(x)$ and $h(x)$ associated to the dark strings as well as the field $R(x)$ associated to the
unbroken U(1) symmetry. The energy per unit length of the dark strings is lowered from
$\mu^{(3,0)}=3$ for $\varepsilon=0$ to $\mu^{(3,0)}=2.99080$ for $\varepsilon=0.1$.

Next, we have investigated the existence of bound states in our system.
Following the definition of the binding energy (\ref{binding}), we present our results
for $\varepsilon \neq 0$ and $\beta_1=\beta_2\equiv \beta$. In the $\varepsilon=0$ limit, $\mu^{(n,m)}_{bin}> 0$ for $\beta >2$,
$\mu^{(n,m)}_{bin} < 0$ for $\beta < 2$ and $\mu^{(n,m)}_{bin} = 0$ for $\beta=2$ (the BPS limit). This is different here as the tables below suggest. In Tables 1,2,3 and 4,5,6 we give the binding energy for $\varepsilon=0.001$ and
$\varepsilon=0.01$, respectively, for different
choices of $\beta$ and $n$ and $m$.

\begin{table}[htb]

\centering{}\begin{tabular}{|c|c|c|c|c|c|}
\hline 
(n,m) & 1 & 2 & 3 & 4 & 5\tabularnewline
\hline
\hline 
1 & -0.0004 & -0.0006 & -0.0008 & -0.0009 & -0.0009\tabularnewline
\hline 
2 & -0.0006 & -0.0011 & -0.0014 & -0.0016 & -0.0017\tabularnewline
\hline 
3 & -0.0008 & -0.0014 & -0.0028 & -0.0022 & -0.0024\tabularnewline
\hline 
4 & -0.0009 & -0.0016 & -0.0022 & -0.0026 & -0.0030\tabularnewline
\hline 
5 & -0.0009 & -0.0017 & -0.0024 & -0.0030 & -0.0034\tabularnewline
\hline
\end{tabular}
\caption{The value of the binding energy $\mu_{bin}^{(n,m)}$ in units of $2\pi$ for $\beta_1=\beta_2=2$, $\varepsilon=0.001$
and different choices of $n$ and $m$.}
\end{table}

\begin{table}[htb]

\centering{}\begin{tabular}{|c|c|c|c|c|c|}
\hline 
(n,m) & 1 & 2 & 3 & 4 & 5\tabularnewline
\hline
\hline 
1 & -0.0005 & 0.0566 & 0.1319 & 0.2169 & 0.3082\tabularnewline
\hline 
2 & 0.0566 & 0.1134 & 0.1886 & 0.2734 & 0.3646\tabularnewline
\hline 
3 & 0.1319 & 0.1886 & 0.2635 & 0.3482 & 0.4393\tabularnewline
\hline 
4 & 0.2169 & 0.2734 & 0.3482 & 0.4328 & 0.5237\tabularnewline
\hline 
5 & 0.3082 & 0.3646 & 0.4393 & 0.5237 & 0.6144\tabularnewline
\hline
\end{tabular}
\caption{The value of the binding energy $\mu_{bin}^{(n,m)}$ in units of $2\pi$ for $\beta_1=\beta_2=3$, $\varepsilon=0.001$
and different choices of $n$ and $m$.}
\end{table}

\begin{table}[htb]
\centering{}\begin{tabular}{|c|c|c|c|c|c|}
\hline 
(n,m) & 1 & 2 & 3 & 4 & 5\tabularnewline
\hline
\hline 
1 & -0.0005 & 0.1049 & 0.2448 & 0.4033 & 0.5738\tabularnewline
\hline 
2 & 0.1049 & 0.2099 & 0.3497 & 0.5081 & 0.6784\tabularnewline
\hline 
3 & 0.2448 & 0.3497 & 0.4893 & 0.6475 & 0.8177\tabularnewline
\hline 
4 & 0.4033 & 0.5081 & 0.6475 & 0.8055 & 0.9755\tabularnewline
\hline 
5 & 0.5738 & 0.6784 & 0.8177 & 0.9755 & 1.1454\tabularnewline
\hline
\end{tabular}
\caption{The value of the binding energy $\mu_{bin}^{(n,m)}$ in units of $2\pi$ for $\beta_1=\beta_2=4$, $\varepsilon=0.001$
and different choices of $n$ and $m$.}
\end{table}

\begin{table}[htb]

\begin{centering}
\begin{tabular}{|c|c|c|c|c|c|}
\hline 
(n,m) & 1 & 2 & 3 & 4 & 5\tabularnewline
\hline
\hline 
1 & -0.0042 & -0.0064 & -0.007 & -0.0085 & -0.0090\tabularnewline
\hline 
2 & -0.0064 & -0.0109 & -0.0138 & -0.0158 & -0.0171\tabularnewline
\hline 
3 & -0.0077 & -0.0138 & -0.0183 & -0.0216 & -0.0240\tabularnewline
\hline 
4 & -0.0085 & -0.0158 & -0.0216 & -0.0262 & -0.0297\tabularnewline
\hline 
5 & -0.0090 & -0.0171 & -0.0240 & -0.0297 & -0.0344\tabularnewline
\hline
\end{tabular}
\caption{ The value of the binding energy $\mu_{bin}^{(n,m)}$ in units of $2\pi$  for $\beta_1=\beta_2=2$, $\varepsilon=0.01$
and different choices of $n$ and $m$.}

\par\end{centering}

\end{table}

\begin{table}[htb]

\centering{}\begin{tabular}{|c|c|c|c|c|c|}
\hline 
(n,m) & 1 & 2 & 3 & 4 & 5\tabularnewline
\hline
\hline 
1 & -0.0046 & 0.0501 & 0.1240 & 0.2081 & 0.2988\tabularnewline
\hline 
2 & 0.0501 & 0.1023 & 0.1743 & 0.2571 & 0.3468\tabularnewline
\hline 
3 & 0.1240 & 0.1743 & 0.2246 & 0.3258 & 0.4142\tabularnewline
\hline 
4 & 0.2081 & 0.2571 & 0.3258 & 0.4055 & 0.4926\tabularnewline
\hline 
5 & 0.2988 & 0.3468 & 0.4142 & 0.4926 & 0.5784\tabularnewline
\hline
\end{tabular}
\caption{The value of the binding energy $\mu_{bin}^{(n,m)}$ in units of $2\pi$ for $\beta_1=\beta_2=3$, $\varepsilon=0.01$
and different choices of $n$ and $m$.}
\end{table}

\begin{table}[htb]

\centering{}\begin{tabular}{|c|c|c|c|c|c|}
\hline 
(n,m) & 1 & 2 & 3 & 4 & 5\tabularnewline
\hline
\hline 
1 & -0.0049 & 0.0978 & 0.2362 & 0.3936 & 0.5634\tabularnewline
\hline 
2 & 0.0978 & 0.1978 & 0.3341 & 0.4900 & 0.6587\tabularnewline
\hline 
3 & 0.2362 & 0.3331 & 0.4684 & 0.6226 & 0.7899\tabularnewline
\hline 
4 & 0.3936 & 0.4900 & 0.6226 & 0.7752 & 0.9410\tabularnewline
\hline 
5 & 0.5634 & 0.6587 & 0.7899 & 0.9410 & 1.1053\tabularnewline
\hline
\end{tabular}
\caption{The value of the binding energy $\mu_{bin}^{(n,m)}$ in units of $2\pi$ for $\beta_1=\beta_2=4$, $\varepsilon=0.01$
and different choices of $n$ and $m$.}
\end{table}

Increasing $\varepsilon$ increases the attractive interaction between the dark string and the comic
string. For $\varepsilon=0$ and $\beta_1=\beta_2=2$, the strings do not interact, while for the same choice
of the $\beta$s and 
$\varepsilon\neq 0$ they form bound states for all values of $n$ and $m$. We observe that
the binding becomes stronger with increasing $n+m$. This is related to the fact that the solutions' mass
increases with increasing winding such that the binding mechanism is more efficient.
Moreover, the binding is stronger for an $(n,n)$ string as compared to an $(n+1,n-1)$ string and the binding
is more effective for an $(n+1,n-1)$ string than for an $(n+2,n-2)$ string etc.
Apparently, dark strings bind stronger to cosmic strings if they have the same winding.

For $\varepsilon=0$ and
$\beta_1=\beta_2=3$ and $\beta_1=\beta_2=4$, respectively, the strings repel since in this case the radius of
the flux tube core is larger than that of the scalar core. Apparently, in this case the attractive
nature of the new interaction ($\varepsilon\neq 0$) is not strong enough to overcome the repulsion for these choices of $\beta_1=\beta_2$, expect in the case $n=m=1$.  Apparently, the attractive interaction
is just about able to overcome the repulsion that two strings would exert on each other for $\epsilon=0$ and
$\beta_i > 2$, $i=1,2$.

\begin{figure}[htbp]
\centering
    \includegraphics[width=13cm]{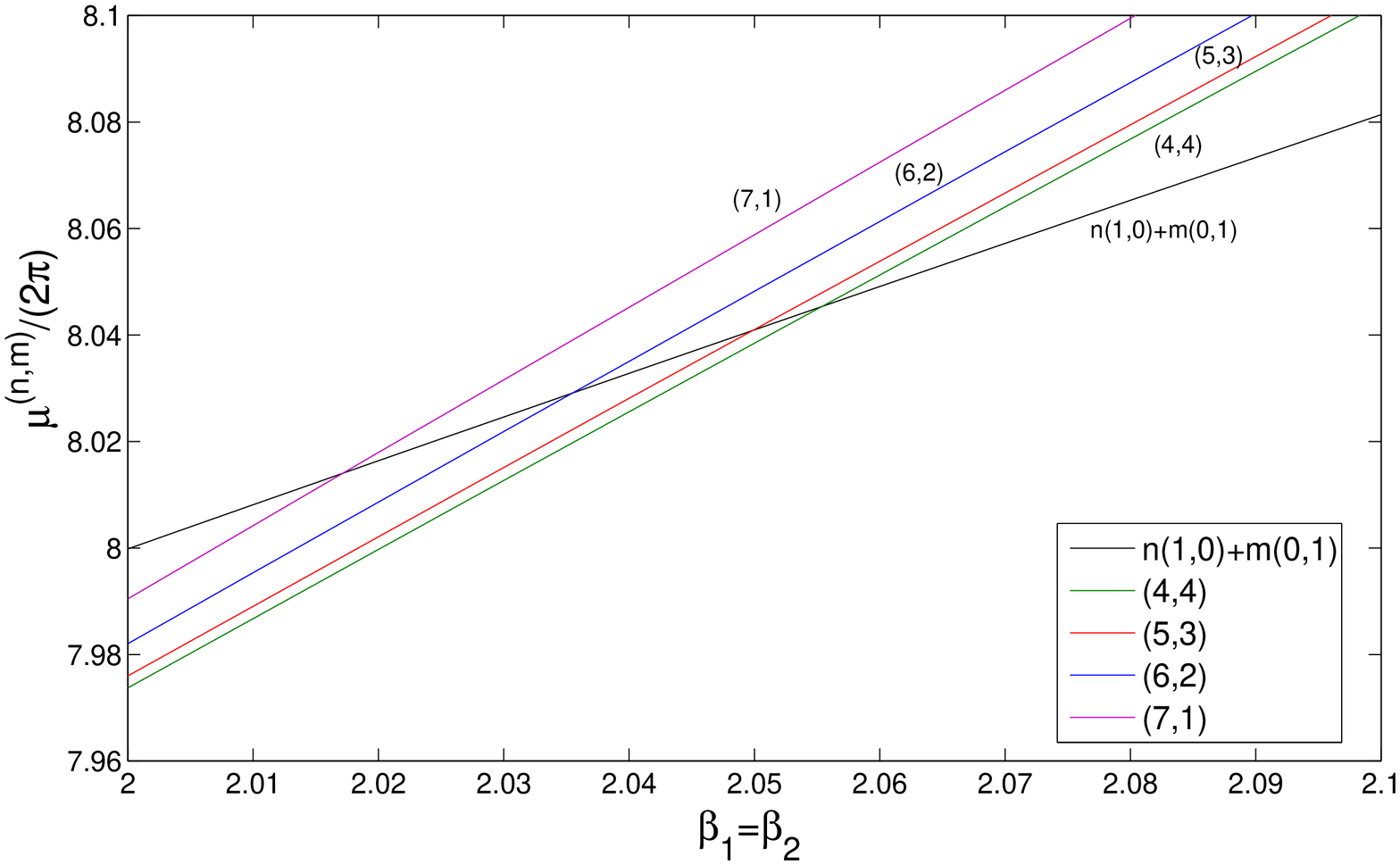} 
\caption{The value of the inertial mass $\mu^{(n,m)}$ in units of $2\pi$ is given in dependence on $\beta_1=\beta_2$ for 
$\epsilon=0.01$ and different choices of $n$ and $m$. Note that $n+m=8$.
For comparison, we also plot $n$ times the mass of the $(1,0)$ solution plus $m$ times the mass
of the $(0,1)$ solution. At the intersection of the $(n,m)$ curves with this latter curve, we have
$\mu_{bin}^{(n,m)}=0$.}
\label{fignew}
\end{figure}

We have next determined the critical value of $\beta_1=\beta_2\equiv \beta=\beta_{cr}$ for which the transition between bound and unbound strings
takes place. For $\varepsilon=0$, this happens at $\beta_{cr}(\epsilon=0)=2$ such that
for $\beta < 2$, the binding energy is negative, while
for $\beta > 2$, the binding energy is positive.
Here, the additional attractive interaction allows us to increase the ratio between Higgs and gauge boson
mass to values larger than $\beta=2$ before the dark string and cosmic string become unbound.
To understand the dependence of $\beta_{cr}$ on $\epsilon$, we have chosen solutions with $n+m$ constant
and have determined the corresponding $\beta_{cr}$ for $\epsilon=0.01$.
Our results are given in Fig.\ref{fignew}.

We plot the mass $\mu^{(n,m)}$ of solutions with $(n,m)=(4+k,4-k)$, $k=0,1,2,3$. For comparison,
we also plot $(4+k)$ times the mass of an $(1,0)$ solution plus $(4-k)$ times the mass of an $(0,1)$ solution.
At the intersection points of this latter curve with the $(4+k,4-k)$ curves, we have $\mu^{(n,m)}_{bin}=0$.
For all solutions, we observe that the mass increases linearly with $\beta_1=\beta_2\equiv \beta$ in the range 
$\beta\in [2:2.1]$ and that the mass of an $(4+k_1,4-k_1)$ solution is
higher than that of a $(4+k_2,4-k_2)$ solution for $k_1 > k_2$. 
Moreover, $\beta_{cr}$, i.e. the value at which the strings become unbound is largest for the solution with $n=4$, $m=4$ and decreases for $k$ increasing. Again, this results from the fact that
dark strings ``bind best'' to cosmic strings which have the same winding.

The magnetic fields (\ref{magnetic}) change as compared to the $\varepsilon=0$ case. We observe that
$\tilde{B}_z(0)$ decreases with increasing $\varepsilon$, while $b_z(0)$ increases with increasing $\varepsilon$.
In addition the core radii of both flux tubes decrease with increasing $\varepsilon$.

\subsection{$\gamma\neq 0$}

\begin{figure}[htbp]
\centering
    \includegraphics[width=13cm]{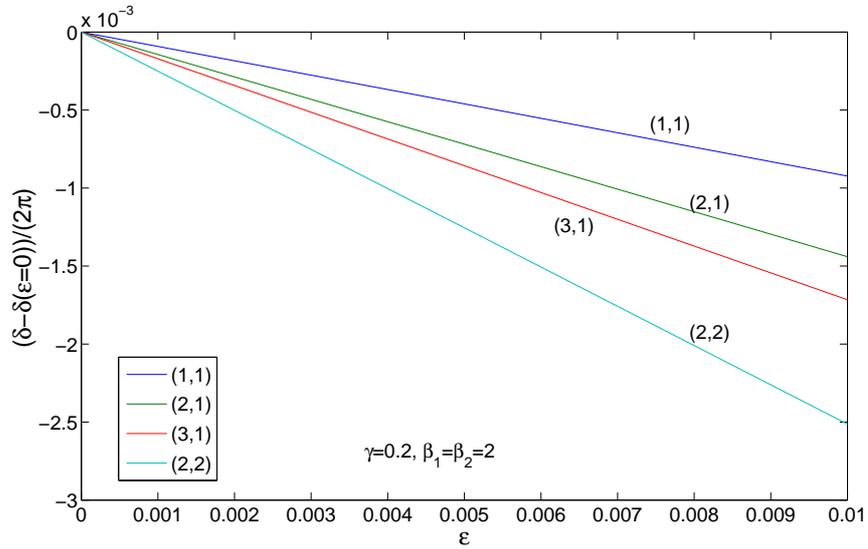} 
\caption{The difference between the deficit angle $\delta/(2\pi)$ and the deficit angle
for $\varepsilon=0$, $\delta(\varepsilon=0)/(2\pi)$ is given as function of $\varepsilon$ for $\gamma=0.2$, $\beta_1=\beta_2=2$ and different choices of $(n,m)$.
 }
\label{fig2}
\end{figure}

We have first studied the dependence of the deficit angle $\delta$ on the interaction parameter
$\varepsilon$. Our results are shown in Fig.\ref{fig2} for $\gamma=0.2$, $\beta_1=\beta_2=2$ and 
different choices of $n$ and $m$. For $\varepsilon=0$, the deficit angle $\delta(\varepsilon =0)$
is given by (\ref{deficitBPS}). Here, we plot the difference between $\delta(\varepsilon=0)$ and
$\delta(\varepsilon\neq 0)$. The deficit angle decreases with increasing $\varepsilon$ and apparently
decreases (approximately) linearly with $\varepsilon$. The larger the sum $n+m$, the stronger $\delta$
decreases with $\varepsilon$, which is related to the fact that the higher $n+m$, the more massive
the solutions are and the more effective is the attractive interaction between the 
dark string and the cosmic string. Moreover, for a fixed sum $n+m$, the deficit angle
decreases stronger for an $(n,n)$ string as compared to an $(n+1, n-1)$ string.
The reason for this is that the binding mechanism is better for a dark string and
a cosmic string with equal winding. This has also been observed before in models describing
interacting cosmic strings when the interaction is mediated via a potential term \cite{saffin,hu}.

Another important feature of cosmic strings in a curved space-time is that
globally regular gravitating solutions exist only
up to a maximal value of the gravitational coupling $\gamma=\gamma_{max}(\varepsilon)$. For $\gamma > \gamma_{max}$, the deficit angle becomes larger than $2\pi$, i.e. the solutions become
singular. For $\varepsilon=0$, these solutions have been denominated ``supermassive'' string solutions \cite{supermassive}. For $\varepsilon=0$ and  $\beta_1=\beta_2=2$ we have $\gamma_{max}(\varepsilon=0)=1/(n+m)$.
For $\varepsilon >0$, the solutions exist on a larger interval of $\gamma$. We find e.g. for $n=1$, $m=1$ and
$\beta_1=\beta_2=2$ that $\gamma_{max}(\varepsilon=0.01)\approx 0.501$, while  $\gamma_{max}(\varepsilon=0.1)\approx 0.515$. This is not surprising since the total energy  per unit length of the solutions decreases
for increasing $\varepsilon$. Since the deficit angle is proportional to the product 
of the gravitational coupling and the energy per unit length, the gravitational coupling can
be increased stronger before $\delta$ becomes equal to $2\pi$.

\begin{figure}[htbp]
\centering
    \includegraphics[width=13cm]{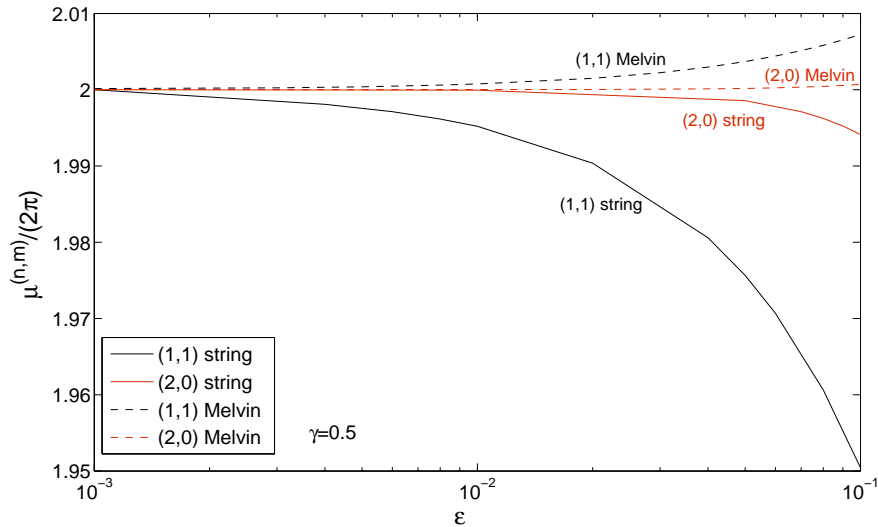} 
\caption{The value of the inertial mass $\mu^{(n,m)}$ of string and Melvin solutions
is shown in dependence on the interaction parameter $\varepsilon$ and $\gamma=0.5$. For $n=m=1$, we have chosen $\beta_1=\beta_2=2$, while
for $n=2$, $m=0$, we have $\beta_1=2$, $\beta_2=0$. }
\label{fig3}
\end{figure}

For $n\neq 0$ and $m=0$, we find the same phenomenon as in flat space-time: dark string solutions can
lower their energy by coupling to the gauge field of an unbroken U(1) symmetry.
In Fig.\ref{fig3}, we compare the masses of string-like solutions for $n=1$, $m=1$ 
and $n=2$, $m=0$, respectively, for $\gamma=0.5$. For $n=1$, $m=1$, we have $\beta_1=\beta_2=2$, while
for $n=2$, $m=0$, we have $\beta_1=2$, $\beta_2=0$. In both cases, the mass of the solution is decreasing
with increasing interaction. It is thus apparent that also in curved space-time,
a dark string can lower its energy when coupling to an unbroken U(1) gauge symmetry.
Moreover, the total energy of the $(1,1)$ system is always
lower than that of the $(2,0)$ system. Hence, the binding is more efficient if a dark string
and a cosmic string couple than if a dark string couples to a pure gauge field.

It has been observed that string-like solutions with a behaviour of the metric functions
$N(x\rightarrow  \infty) \rightarrow c_1$ and $L(x\rightarrow \infty) \rightarrow c_2 x  + c_3$
have a ``shadow solution'' in the form of so-called Melvin solutions \cite{clv,bl}.
In contrast to string-like solutions, Melvin solutions have no flat space-time counterparts and their
metric functions behave as $N(x\rightarrow \infty)\rightarrow  a_1 x^{2/3}$ and $L(x\rightarrow \infty)\rightarrow
a_2 x^{-1/3}$, where $a_1$ and $a_2$ are constants.
These solutions are also present in systems where cosmic strings interact via a potential
term \cite{hu} and we also find them here.
In Fig.\ref{fig3}, we give the inertial mass of these solutions for two different choices of $n$ and $m$.
When comparing to the string solutions, it is apparent that the mass of the Melvin solutions
is higher for all choices of the interaction parameter. Moreover, in contrast to string solutions,
the inertial mass of the Melvin solutions increases for increasing interaction.
This has also been observed in \cite{hu}, where two cosmic strings interact via a potential.
The stronger the attractive interaction between the strings, the higher the inertial mass of the Melvin 
solutions. In addition, the mass increases stronger for an $(1,1)$ system of dark strings than
for an $n=2$ dark string coupled to the gauge field of the unbroken U(1) symmetry.
This is related to the observation that the binding is stronger for the $(1,1)$ system than for the $(2,0)$ system.

\section{Conclusions and Outlook}
 In this paper, we have studied the interaction of a dark string with a cosmic string where the
interaction is mediated by a coupling term between the field strength tensors of the respective solutions.
This type of interaction is motivated by recent models describing the dark matter sector.
We observe that a BPS bound exists if the dark string is identical to the cosmic string.
In fact, we find that the attractive interaction between the two strings is most efficient
in this particular case. In addition, the attractive interaction allows for
dark-cosmic strings to exist for larger values of the Higgs to gauge boson ratio.
The deficit angle associated to the strings decreases for increasing interaction and
globally regular string space-times exist for higher values of the gravitational
coupling as compared to the non-interacting case.
We also find dark string solutions that can lower their energy by coupling to a U(1)
field through the attractive interaction term.

The formation of bound states is of interest for the study of the evolution of
string networks. For p-q-strings it has been observed \cite{wyman} that the formation
of bound states leads to an energy loss mechanism that is important 
in the evolution of string networks towards the scaling regime.
If the dark matter sector has indeed dark strings solutions then
standard cosmic strings could loose energy by coupling to these strings.

It would also be interesting to understand the influence of the type of interaction studied here on superconducting
strings \cite{witten2}. In its simplest version this would be the coupling of a U(1)$\times$ U(1) model (describing
the superconducting string with bosonic charge carriers) coupled to a U(1) model describing the
dark string. The field strength tensor of the dark string could then either be coupled
to the field strength tensor associated to the broken U(1) symmetry (similar to what has been studied in this
paper) or alternatively to the field strength tensor of the unbroken U(1) symmetry describing the carrier field.

\end{document}